\documentclass[conference]{IEEEtran}
\IEEEoverridecommandlockouts
\usepackage{cite}
\usepackage{amsmath,amssymb,amsfonts}
\usepackage{algorithmic}
\usepackage{graphicx}
\usepackage{textcomp}
\usepackage{xcolor}
\usepackage{multirow}
\usepackage{threeparttable}
\usepackage{booktabs}
\usepackage{amssymb}
\usepackage{epstopdf}
\usepackage{CJKutf8}
\def\BibTeX{{\rm B\kern-.05em{\sc i\kern-.025em b}\kern-.08em
    T\kern-.1667em\lower.7ex\hbox{E}\kern-.125emX}}
\begin{document}

\title{Multi-Scale Temporal Convolution Network for Classroom Voice Detection\\
\thanks{This work was supported by National Key R$\&$D Program of China, under Grant No. 2020AAA0104500. The corresponding author is Lu Ma.}
}

\author{\IEEEauthorblockN{Lu Ma, Xintian Wang, Song Yang, Yaguang Gong, Zhongqin Wu}
\IEEEauthorblockA{TAL Education Group \\
Beijing, China \\
\{malu6,wangxintain,yangsong1,gongyaguang,wuzhongqin\}@tal.com}}

\maketitle

\begin{abstract}
Teaching with the cooperation of expert teacher and assistant teacher, which is the so-called ``double-teachers classroom'', i.e., the course is giving by the expert online and presented through projection screen at the classroom, and the teacher at the classroom performs as an assistant for guiding the students in learning, is becoming more prevalent in today's teaching method for K-12 education. For monitoring the teaching quality, a microphone clipped on the assistant's neckline is always used for voice recording, then fed to the downstream tasks of automatic speech recognition (ASR) and neural language processing (NLP). However, besides its voice, there would be some other interfering voices, including the expert's one and the student's one. Here, we propose to extract the assistant' voices from the perspective of sound event detection, i.e., the voices are classified into four categories, namely the expert, the teacher, the mixture of them, and the background. To make frame-level identification, which is important for grabbing sensitive words for the downstream tasks, a multi-scale temporal convolution neural network is constructed with stacked dilated convolutions for considering both local and global properties. These features are concatenated and fed to a classification network constructed by three linear layers. The framework is evaluated on simulated data and real-world recordings, giving considerable performance in terms of precision and recall, compared with some classical classification methods.
\end{abstract}


\section{Introduction}
\begin{figure}[htb]
\centering
\includegraphics[width=0.43\textwidth]{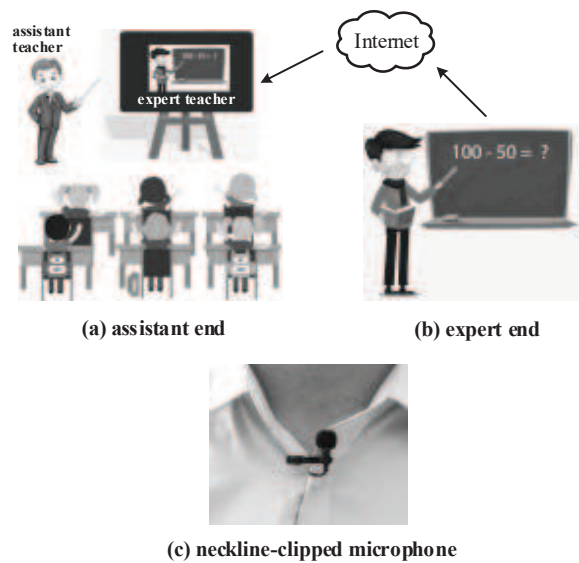}
\caption{The block diagram of the network structure.} \label{fig0}
\end{figure}

Nowadays, a new teaching mode named ``double-teachers classroom'' is arising. It is carried out with the cooperation of an expert teacher online and an assistant teacher in the classroom. The experts are teachers with excellent teaching skills, locating at the distance, and present through projection screen in the classrooms. The teacher in the classroom performs as an assistant, guiding the students to learn the courses and maintain the classroom discipline. The scenario is depicted in Fig.~\ref{fig0}(a)(b).
It is a unique instruction mode designed to share high quality teaching resources with students in remote areas as a means of providing education equality~\cite{ref_double_teacher}.

To monitor the teaching quality, the voice of the assistant teacher is always recorded by a microphone clipped on the neckline as is shown in Fig.~\ref{fig0}(c). The neckline-clipped microphone is always used because it is a commercial shelf product (COTS) and is convenient to obtain. The detected assistant teacher's voice will be fed to the downstream tasks of automatic speech recognition (ASR) and neural language processing (NLP) for teaching monitoring. However, besides the assistant teacher's own voice, there would be some other voices from the surroundings, including the expert's one, the student's one and the background.

An intuitive way for tackling this problem is taking sound separation methods into account, such as deep clustering (DPCL)~\cite{ref_dpcl}, Conv-Tasnet~\cite{{ref_convtasnet}} and DPRNN~\cite{ref_dprnn}. However, they are mostly constructed at utterance level with segmental length of 4 second (for example), where the voice overlapping is supposed to occur within this segment. And, the number of speakers within this utterance is constant. However, in our scenario, the voice duration per source would last much long, i.e., only one voice exists within this segment, resulting in a varying number of speakers. Moreover, the binary mask estimated from DPCL method would damage the speech spectrogram, bringing harmful to the downstream tasks, such as ASR. These will lead to poor performance in our situation for current separation framework. Another way is from the perspective of semantic analysis with NLP technology using the transcriptions from ASR, such as~\cite{ref_cad2}\cite{ref_cad3}.

From observations, since the expert teacher's voice is played by the screen far away from the microphone, it has worse reverberation than that of the assistant teacher, and the voice from individual student would be too low to be recorded. Only when they yell together at a high level, merely in some situations of answering questions, could the voice be recorded. Otherwise, it performs like babble noise in most situations. In addition, the mixture voice would also arise at a relatively low frequency. Since the voices present different characteristics, we propose to detect the voice from the perspective of sound event detection (SED)~\cite{ref_sed}, i.e., the voices are classified into four categories, namely the expert, the assistant, the mixture of them, and the background. Our scenario is more complicated than~\cite{ref_cad4}\cite{ref_cad1} where three types of plain voices are required to be classified, i.e., no-voice, single-voice, and multi-voice. Moreover, it is notable that frame-level identification of voice detection is required for teaching monitoring, especially for grabbing sensitive words for downstream tasks of ASR and NLP, to do some analysis such as whether it is a question, or whether it is a praise.

Therefore, inspired by~\cite{ref_convtasnet}\cite{ref_tcn}, we constructed a multi-scale temporal convolution neural network (MSTCN) for classification. The multi-scale features are extracted with stacked dilated temporal convolutions for considering both local and global properties. These features from different layers are concatenated and fed to a classification network constructed by three linear layers. The model is first trained with manufactured data, and then evaluated on the simulated testing data and real-world recordings. Experiments reveal that higher performance in terms of precision and recall can be achieved compared with some classical methods\cite{ref_mobilenet}\cite{ref_fcn}\cite{ref_dcase}.

The remainder of this paper is organized as follows. Section 2 provides the details of our proposed network structure. Section 3 presents our experimental results. And finally, the summarization and discussions are given in Section 4.

\begin{figure*}[htb]
\centering
\includegraphics[width=0.8\textwidth]{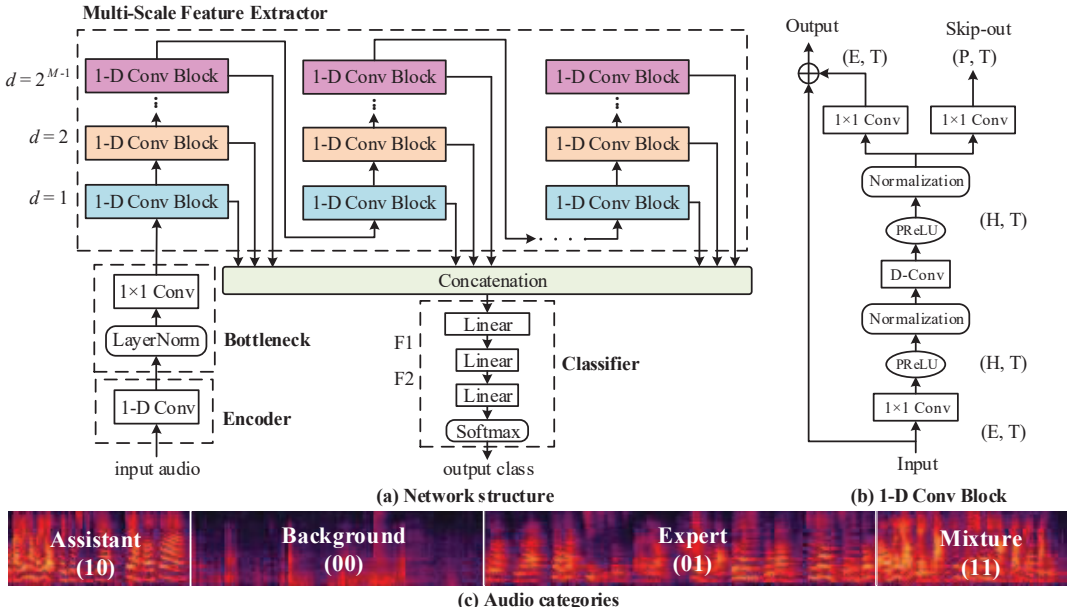}
\caption{The block diagram of the network structure.} \label{fig1}
\end{figure*}

\section{Multi-Scale Temporal Convolution Network}

The network structure is depicted in Fig.~\ref{fig1}(a), which is constituted by an encoder, a bottleneck, a multi-scale feature extractor, and a classifier.

\subsection{Encoder} \label{sect_encoder}
It is realized by a 1--D convolution (1--$D$ $Conv$), performing like short-time Fourier transform~\cite{ref_stft} with high-resolution frequency decomposition. The input audio is divided into overlapping segments of length $L$ samples. It can be represented by $x_{k} \in \mathbb{R}^{1 \times L}$, where $k=1, \ldots, {T}$ denotes the segment index and ${T}$ denotes the total number of segments. $x_{k}$ is transformed into a $N$-dimensional representation by a 1--D convolution operation $\mathbf{w} \in \mathbb{R}^{1 \times N}$. It can be formulated by a matrix multiplication as
\begin{equation}
{\mathbf{w}}=\mathcal{H}(\mathrm{\mathbf{x}} \mathbf{U})
\end{equation}
where $\mathbf{U} \in \mathbb{R}^{N \times L}$ contains $N$ vectors (encoder basis functions) with length $L$ for each, $\mathcal{H}(\cdot)$ is the rectified linear unit (ReLU) function~\cite{ref_relu} to ensure non-negative of the representation.

\subsection{Bottleneck} \label{sect_bottle}

The layer normalization used here is to ensure that the following extracted feature is invariant to the input scaling. The 1--D convolution block is used to compress the number of input channels~\cite{ref_bottleneck} from $N$ to $E$, obtaining higher dimensional features.

\subsection{Multi-Scale Feature Extractor}

Motivated by the Conv-Tasnet for sound source separation~\cite{ref_convtasnet}, the stacked 1--D dilated convolutional blocks~\cite{ref_wavenet} are used for multi-scale feature extraction. Each layer consists of 1--D convolutional blocks with increasing dilation factors that increase exponentially to ensure a sufficiently large temporal context window to take advantage of the long-range dependencies of the audio signal. It is depicted in Fig.~\ref{fig1}(b) with different colors. $M$ convolutional blocks with dilation factors $d = 1,2,4, \ldots, 2^{M-1}$ are repeated $R$ times for extracting multi-scale features. The input to each block is zero padded accordingly to ensure the same the output length as the input. For each 1--D convolutional block, a residual path and a skip-connection path are applied: the residual path of a block serves as the input to the next block, and the skip-connection paths for all blocks are concatenated and used for classification. Since residual path is used for obtaining deeper network, the number of channels of the residual path is same as the output channel of the bottleneck, denoted by $E$. The number of channel of the skip-connection path can be different with that of the residual path, but they should be same among different layers, and can be denoted by $P$. As is the same with the Conv-Tasnet, depthwise separable convolution is used to replace standard convolution in each convolutional block to decrease the number of parameters, which is performed with a depthwise convolution ($D$-$Conv(\cdot)$) followed by a pointwise convolution ($1$$\times $$1$-$Conv(\cdot)$)~\cite{ref_depthwise}\cite{ref_depthwise1}. The parametric rectified linear unit (PReLU) activation function and a normalization operation are performed after both first $1$$\times $$1$-$Conv(\cdot)$ and $D$-$Conv(\cdot)$ blocks respectively. The normalization can be chosen as global layer normalization (gLN) for noncausal configuration and cumulative layer normalization (cLN) for causal configuration~\cite{ref_convtasnet}\cite{ref_layernorm}.

The feature extractor consists of $Q$ layers and each layer is denoted by $\mathbf{e}^{(q)} \in \mathbb{R}^{F_{q} \times T_{q}}$, where ${F_{q}}$ is the length of convolutional filter in the $q$-th layer and $T_{q}$ is the length of corresponding time steps. In this paper, we use the same length of convolutional filter for all layers, i.e., $F_{q}=P$, and all the layers are with the same time steps, i.e., $T_{q}={T}$. Here, $P$ and ${T}$ are described in Subsections \ref{sect_bottle} and \ref{sect_encoder}. Thus, we define the collection of filters in each layer as $W = \left\{W^{(i)}\right\}_{i=1}^{P}$ for $W^{(i)} \in \mathbb{R}^{d \times P}$ with a corresponding bias vector $b \in \mathbb{R}^P$, and $d$ is the dilation factors. Given the signal from the previous layer $\mathbf{e}^{(q-1)}$, we compute $\mathbf{e}^{(q)}$ with

\begin{equation}
\mathbf{e}^{(q)}=f\left(W * \mathbf{e}^{(q-1)}+b\right)
\end{equation}
where $f(\cdot)$ is the the activation function, $*$ is the convolution
operator.

The detailed description of symbols for the network are described in Table~\ref{tab:configuration}.

\begin{table}[htb]
\centering
\caption{Network configurations}
\label{tab:configuration}
\begin{tabular}{c|c}
\hline
\textbf{Symbol} & \textbf{Description} \\ \hline \hline
$T$ & Number of segments \\ \hline
$N$ & Number of filters in encoder and decoder \\ \hline
$L$ & Length of the filters (in samples) \\ \hline
\multirow{2}{*}{$E$} & Number of channels in bottleneck and \\
  & the residual paths' 1--D conv block \\ \hline
\multirow{2}{*}{$P$} & Number of channels in skip-connection \\
                    & paths' of 1--D conv block \\ \hline
$H$ & Number of channels in convolutional blocks \\ \hline
$K$ & Kernel size in convolutional blocks \\ \hline
$X$ & Number of convolutional blocks in each repeat \\ \hline
$R$ & Number of repeats \\ \hline
$F$ & Hidden size of linear layers, $F_1$, $F_2$ \\ \hline
$C$ & Number of classes for classification \\ \hline
\end{tabular}
\end{table}

\subsection{Classifier}\label{AA}
Linear layers with input size of $M$$\times R$$\times$$ P$ and output size of 2, followed by a softmax activation function~\cite{{ref_softmax}} is used for classification. Here, three layers are used, with the first and second hidden size as $F_1$ and $F_2$. Here, two output classes as is shown in Fig.~\ref{fig1}(c) are used, i.e., one class represents the assistant teacher and the other represents the expert teacher. A threshold denoted by $G_{th}$ is used for each class for voice detection. Therefore, four categories can be obtained, i.e, only background noise (denoted by ``00''), only the expert's voice (denoted by ``01''), only the assistant's voice (denoted by ``10''), and the mixture of the expert and the assistant (denoted by ``11'').

\section{Experiments}

\begin{table*}[htbp]
\label{tab:comparisons}
\normalsize
\begin{center}
\renewcommand\arraystretch{1.2}
\begin{threeparttable}
\begin{tabular}{cccccccccc} \toprule
\caption{Experimental results on the datasets}
    \multirow{2}{*}{Dataset} & \multirow{2}{*}{Methods} & \multicolumn{2}{c}{Background} & \multicolumn{2}{c}{Expert} & \multicolumn{2}{c}{Assistant} & \multicolumn{2}{c}{Mixture} \\ \cmidrule(r){3-4} \cmidrule(r){5-6} \cmidrule(r){7-8} \cmidrule(r){9-10}
    & & Rec. & Pre. & Rec. & Pre. & Rec. & Pre. & Rec. & Pre. \\ \hline
    \multirow{4}{*}{Simulations} & MobileNet & 99.73 & 99.91 & 99.86 & 99.90 & 97.91 & 98.24  & 98.15 & 97.87  \\
& FCNN & 99.86 & 99.90 & 99.91 & 99.81 & 98.62 & 97.97  & 97.91  & 98.61  \\
& TCN & 99.66 & 99.99 & 99.61  & 99.32  & 97.74  & 98.67  & 98.37  & 97.39 \\
& MSTCN & 99.93 & 99.99 & 99.93  & 99.41  & 98.74  & 98.75  & 98.67  & 98.79  \\ \hline
    \multirow{4}{*}{Recordings} & MobileNet & 72.17 & 96.59 & 98.85 & 45.11 & 63.06 & 96.71  & 44.34 & 60.92  \\
& FCNN & 78.61  & 73.53  & 75.11  & 84.14  & 64.71  & 99.33 & 43.22  & 55.53  \\
& TCN & 85.73 & 92.33  & 93.38  & 83.36  & 73.26  & 97.66   & 45.23 & 70.32 \\
& MSTCN & 87.91 & 93.67  & 95.32  & 85.73  & 85.61  & 98.53   & 51.35 & 98.91  \\
    \bottomrule
\end{tabular}
\end{threeparttable}
\end{center}
\end{table*}

\subsection{Data Preparation}
For simulations, audios from two speakers randomly selected from the $Librispeech$ dataset~\cite{ref_librispeech} are considered as the expert and the assistant. The power ratio between the assistant and the expert is chosen between 3dB and 12dB with 1dB each step. Audio randomly chosen from the $wham$ noise dataset~\cite{ref_wham} is considered as the background. The signal-to-noise ratio of the expert with respect to the noise is chosen between 5dB and 15dB with 1dB each step. The reverberation time is selected from 0.4 second (s) to 0.9s with 0.1s each step, and the room size is fixed as $(x,y,z) = [5.0$m$, 6.0$m$, 3.5$m$]$. The expert is fixed at $[2.5$m$, 0.5$m$, 2.0$m$]$. The $y$-axis and $z$-axis for the assistant are constant as $1.0$m and $1.6$m, with the $x$-axis position changing from $0.5$m to $4.5$m with $0.5$m each step. The microphone is $0.01$m lower than the assistant in $z$-axis. The audio is convoluted with the corresponding room impulse responses (RIRs) generated by $gpuRIR$~\cite{ref_gpurir}.

According to the aforementioned configurations, four type of utterances are generated, i.e., pure background noise, expert's voice, assistant's voice, the mixture of the expert and the assistant. Each utterance is 3 second long. Each time, these four type of utterances are concatenated in time axis with random order to obtain a audio file with 12s length for model training. It is similar to that shown in Fig.~\ref{fig1}(c). In total, $4$$\times$$10^4$, $1$$\times$$10^4$ and $1$$\times$$10^3$ samples are generated for training, validation and testing, each with 12s duration. In the training and validation, the processing window is 3s length and each time a new frame is feed into the window from the most right side and the old frame at the most left side slides out the window. Each time, the classification result within this window is considered as tag of the middle frame of the window.

For real-world recordings, 9 audio files with 10min length for each are used for testing using the model trained on the aforementioned simulation data. No recording data are used for model training. These 9 audio files are collected from 9 different classrooms. The room size is same as the simulation, i.e., $(x,y,z) = [5.0$m$, 6.0$m$, 3.5$m$]$. A microphone clipped on the neckline of the assistant teacher is used for voice recording. The height of the assistant teachers are ranging from 1.55m to 1.65m. And the assistant teacher locates about 1m distance to the screen where the expert teacher's voice is played out.

\subsection{Model Configuration}
The threshold for voice detection is $G_{th} = 0.5$. cLN is used for normalization. Cross-entropy is selected as loss function. $Adm$ algorithm is used for training with an exponential learning rate decaying strategy, where the learning rate starts at $10^{-5}$ and ends at $10^{-8}$. The total number of epochs is set to be 150, and the criteria for early stopping is no decrease in the loss function on validation set for 10 epochs. The other parameters listed in Table~\ref{tab:configuration} are configured as: $L=160$, $N=512$, $E=128$, $H=512$, $K=3$, $R=3$, $M=8$, $P=128$, $F_1=2048$, $F_2=2048$, $C=2$.

\subsection{Results}
Three classical baselines are used for comparisons, i.e., the MobileNet~\cite{ref_mobilenet}, the FCNN~\cite{ref_fcn}, and the TCN~\cite{ref_tcn}. The MobileNet and the FCNN are configured as same with~\cite{ref_dcase}. For the baseline of the TCN, it's structure is the same with our model, except that the multi-scale feature extractor of our model is replaced by the TCN. For testing of the simulations and the real-word recordings, the processing window is configured as 3 second long in the experiments.

As is shown in Table~\ref{tab:comparisons}, for simulations, our model has higher performance compared with the baselines, in terms of the recall (Rec.) and the precision (Pre.) which are denoted by percentage ($\%$).
For real-world recordings, our model has the best precision performance for the assistant identification compared with other methods. Though, the expert's recall of our model is slightly lower than that of the MobileNet, it gains a higher precision, and outperforms other models. It is notable that, our method outperforms the TCN method in all aspects. This is because that in our method, multi-scale features are utilized instead of only the last layer used in the TCN method. It also can be found that, our method has the best performance both in recall and precision for the mixture classification. Since the model is trained on simulation data and then tested on real-world recordings, it reveals good generalization and fine identification abilities. It is important for frame-level or word-level identification performance in education to grab sensitive words for teaching monitoring, such as whether it is a question, or whether it is a praise. As can be seen from Table~\ref{tab:comparisons}, our model have relatively higher performance considering precision and recall on average.

\section{Conclusions}
We present a multi-scale temporal convolution neural network for voice detection in double-teachers classroom where a microphone clipped on the neckline is used for voice recording. Experiments demonstrate considerable performance in terms of frame-level detection and generalization ability of our framework. The detected assistant teacher's voice will be fed to the downstream tasks such as ASR and NLP for monitoring the teaching quality. In the future, we would like to design multi-model schemes by combining audio, video and transcriptions for voice detection.

\vspace{12pt}

\end{document}